# Monitoring of the formation of SrMoO$_4$ intergrain tunneling barriers in Sr$_2$FeMoO$_{6-\delta}$

Gunnar Suchaneck[a,*], Nikolay Kalanda[b], Evgenij Artsiukh[b], Marta Yarmolich[b], Nikolai A. Sobolev[c]

[a] TU Dresden, Solid State Electronics Laboratory, 01062 Dresden, Germany

[b] SSPA "Scientific-Practical Materials Research Centre of NAS of Belarus", Cryogenic research division, 220072 Minsk, Belarus

[c] i3N, Departamento de Física, Universidade de Aveiro, 3810-193, Aveiro, Portugal

[*] Corresponding author's e-mail: gunnar.suchaneck@mailbox.tu-dresden.de

*Abstract*

This work is a contribution to the understanding of the electrical resistivity in strontium ferromolybdate (Sr$_2$FeMoO$_{6-\delta}$, SFMO) ceramics. It demonstrates that an appropriate thermal treatment leads to the formation of dielectric SrMoO$_4$ shells at the surface of SFMO nanograins. In samples without SrMoO$_4$ shells, the sign of the temperature coefficient of resistance changes with increasing temperature from negative at very low temperature to positive at higher temperatures. Samples exhibiting a negative temperature coefficient contain SrMoO$_4$ shells and demonstrate a behavior of the resistivity that can be described in terms of the fluctuation-induced tunneling model, and near room temperature the conductivity mechanism converts to a variable-range hopping one. The results of this work serve as a starting point for the understanding of the low-field magnetoresistance which is very promising for spintronic device application.

# 1. Introduction

Strontium ferromolybdate ($Sr_2FeMoO_{6-\delta}$, SFMO) double perovskites are promising candidates for magnetic electrode materials for room-temperature spintronic applications, because they present a half-metallic character (with theoretically 100% electron spin polarization), a high Curie temperature ($T_C$) of about 415 K (ferrimagnets should be operated in their ordered magnetic state below $T_C$), and a low-field magnetoresistance [1]. The main reasons for the still missing wide application of SFMO is the low reproducibility of its electrical and magnetic properties originating in the formation of point defects and grain boundaries with a composition different from the bulk of the grains, as well as its aging in contact with air and moisture.

The electrical conductivity of SFMO in the absence of magnetic field shows different behavior as a function of the synthesis conditions and microstructure:

(i) A semiconductor-like behavior with a negative temperature coefficient of resistivity, $d\rho/dT < 0$, is observed up to about room temperature in ceramics sintered at 1200°C in a $H_2/Ar$ atmosphere [1,2]. Note that in this case the electrical transport may occur also by electron hopping between localized states.

(ii) A continuous decrease of the resistivity with increasing temperature in pressed powders [3] and nanograined SFMO samples [4] is described by a two-channel conductivity model where one of the channels is the spin-dependent intergranular tunneling with $d\rho/dT < 0$ dominating at low temperatures. The other channel was attributed to hopping through localized states with $d\rho/dT > 0$ being the conductivity channel at high temperatures.

(iii) A nearly linear decrease of the resistivity from liquid He temperatures up to room temperature in cold-pressed and sintered ceramics is attributed to the fluctuation-induced intergrain tunneling [5–7].

(iv) A semiconductor-like behavior at low temperatures is followed by an upturn in resistivity to metallic behavior, $d\rho/dT > 0$, at higher temperatures for ceramics sintered at 1200°C in highly reducing atmospheres [8] and at 1150°C in vacuum [9]. The origin of the low-temperature upturn in resistivity is not well understood. It might be attributed to a) inelastic scattering of electrons by impurity ions in impure metals [10], b) two parallel conductivity channels - thermally activated hopping and excitation to the mobility edge [11,12], or c) two spin channels in half-metallic double perovskites connected in parallel, where the spin-down channel is metallic (with a non-zero density of states at the Fermi level), and the spin-up channel, with a gap in the band structure, behaves like a semiconductor [13]. The metallic behavior is observed up to the Curie temperature, followed by a semiconducting behavior [9,14] up to the appearance of an insulator-metal transition at 590 K yielding a minimum in resistivity at this temperature [9].

SFMO samples obtained by solid-state reactions using one and the same procedure were insulating, metallic or in an intermediate state in dependence on the precursors and annealing temperature and time [15]. The sintering of sol-gel fabricated, nanograined SFMO ceramics at a low temperature or for a short time at a high temperature was not enough to eliminate the influence of grain boundaries [16]. On the other hand, with increasing post-oxidation of grain boundaries, the resistivity minimum disappears, and the behavior becomes semiconductor-like in the whole temperature range [8]. By soaking SFMO ceramics in a glycerin/water mixture, $SrMoO_4$ precipitates form at grain boundaries causing the low-field magnetoresistance interesting for practical applications [17]. Here the resistivity also increases, and the resistivity minimum as well as a metallic behavior at higher temperatures disappear. A similar change of the conductivity behavior from upturn to semiconductor-like, connected with the formation of $SrMoO_4$ shells at grain boundaries, was obtained in fine-grained SFMO ceramics after a heat treatment for 5 h at $T = 430°C$ in an Ar atmosphere [18]. These ceramics were fabricated using nanograined powders synthesized by the citrate-gel technique. They were finally sintered in a 5%$H_2$/Ar flow at 1000°C for 4 h. After a low-temperature annealing, $SrMoO_4$-related reflexes appeared in the x-ray diffraction (XRD) pattern indicating that the content of this phase did not exceed 7.5%. Additionally, a transition of the valence state of molybdenum cations at grain surfaces ($Mo^{5+} \rightarrow Mo^{6+}$) was proved. The thickness of the dielectric $SrMoO_4$ surface layer, playing the role of a potential barrier between metallic grains, was estimated as 10 nm, whereas the average SFMO grain size decreased down to 50 nm. However, no other barrier characteristics were communicated [18].

In this work we show that a successive oxidation of the SFMO grain boundaries up to the formation of thin insulating $SrMoO_4$ surface layers suppresses the metallic conductivity owing to the appearance of intergrain fluctuation-induced tunneling. Thus, this study is an important first step in solving the puzzle of the impact of synthesis conditions and microstructure on the electrical conductivity of SFMO ceramics. Moreover, such dielectric shells, acting as tunnel barriers around the nanoparticles in SFMO layers, are promising candidates for applications in innovative spintronic devices exhibiting the so-called low-field magnetoresistance [19].

## 2. The fluctuation-induced tunneling model

The fluctuation-induced tunneling (FIT) model describes conducting grains separated by energy barriers subjected to large thermal fluctuations, e.g. carbon-polyvinylchloride composites, consisting of aggregates of carbon spheres [20]. The tunneling occurs between large metallic grains across insulating barriers with a width $w$ and area $A$. It is specified by two parameters: (i) The temperature $T_1$ characterizing the electrostatic energy of a parabolic potential barrier

$$kT_1 = \frac{A \cdot w \cdot \varepsilon_o E_0^2}{2}, \qquad (2.1)$$

where $k$ is the Boltzmann constant and the characteristic field $E_0$ is determined by the barrier height $V_0$:

$$E_0 = \frac{4V_0}{e \cdot w}, \qquad (2.2)$$

(ii) The temperature $T_0$ representing $T_1$ divided by the tunneling constant

$$T_0 = T_1 \cdot \left(\frac{\pi \chi w}{2}\right)^{-1}, \qquad (2.3)$$

with the reciprocal localization length of the wave function

$$\chi = \sqrt{\frac{2m_e V_0}{\hbar^2}}, \qquad (2.4)$$

where $m_e$ is the electron mass. The resulting resistivity in this model is then given by

$$\rho(T) = \rho_0 \exp\left(\frac{T_1}{T_0 + T}\right) \qquad (2.5)$$

The thermal fluctuations reduce both the barrier's height and width. Thus, the conductivity at the temperature $T = T_0$ drops to 1/2 of its extrapolated value at $T = 0$ K. For $T \gg T_0$, conductivity with an activation energy $E_a = kT_1$ appears. The FIT model was recently applied to inter-grain tunneling in polycrystalline $Sr_2CrMoO_6$ and $Sr_2FeMoO_6$ ceramics [6], in half-metallic double-perovskites $Sr_2BB'O_6$ (BB'– FeMo, FeRe, CrMo, CrW, CrRe) ceramics [7] and in $Ba_2FeMoO_6$ thin films [21].

## 3. Experimental

The citrate-gel technique was used for a synthesis of SFMO nanopowders using ultra-high purity $Sr(NO_3)_2$, $Fe(NO_3)_3 \cdot 9H_2O$, $(NH_4)_6Mo_7O_{24}$ and citric acid monohydrate $C_6H_8O_7H_2O$ as initial reagents. To obtain a colloidal sol, aqueous solutions of $Sr(NO_3)_2$ and $Fe(NO_3)_3 \cdot 9H_2O$ were mixed in a molar ratio of Sr/Fe = 2:1. Citric acid was added to the solution in a molar ratio of citric acid/Fe = 6.5:1. After that, an aqueous solution of $(NH_4)_6Mo_7O_{24}$ was added to the solution of strontium and iron nitrates in a molar ratio Mo/Fe = 1:1. Then, ethylenediamine was added upon constant stirring by means of an IKA C-MAG HS7 magnetic stirrer, until the *pH* of the solution reached 4. Thereafter, the substance was dried at a temperature of 80°C. The resulting precipitate was placed in a furnace at a temperature of 100°C, followed by heating at a rate of 0.4°C/min up to 200°C, a dwell time of 18 hours, and a cooling-down with the time constant of the furnace. The result was a solid foam which was crushed and then subjected to heat treatment at 500°C in an oxygen atmosphere under the pressure $p(O_2) = 0.21 \cdot 10^5$ Pa for 10 h. The final SFMO synthesis was carried out in a reducing ambient of a 5%$H_2$/Ar gas mixture at 900°C for 4 hours in several stages. Single-phase SFMO powders were pressed into tablets with a diameter of 10 mm and a thickness of 3 mm under a pressure of 4 GPa at 530°C for 1 min. Dielectric shells were formed on these samples on the surface of the SFMO grains by annealing in an Ar flow with a rate of 11 sccm and for different times at 530°C. Details of sample fabrication were already described elsewhere [18,22].

The phase composition of solid-phase synthesis products, as well as of the SFMO samples, was determined using XRD phase analysis on a DRON-3 apparatus in $CuK_\alpha$ radiation by means of the ICSD-PDF2 database (Release 2000) and PowderCell software. The microstructure of the samples was analyzed using a JEOL JSM 6360 scanning electron microscope (SEM).

X-ray photoelectron spectroscopy (XPS) measurements were carried out by means of a PHI 5600 (Physical Electronics) spectrometer equipped with a monochromatic AlKα (1486.6 eV) source in a vacuum of $10^{-7}$ Pa. The analysed surface amounted to 800 μm$^2$, the angle between the analyser and the sample surface was 45º. The binding energy was calibrated against the C(1s) peak at 284.9 eV related to the surface contamination of the sample, in order to correct charging effects. Peak fitting was performed using the PHI Multipack 9.3 software.

The magnetic and electrical transport properties of the samples were studied in the temperature range from 4.2–600 K in a constant magnetic field in a universal Liquid Helium Free High Field Measurement System (Cryogenic Ltd.).

## 4. Results and discussion

XRD patterns (Fig. 1) revealed a pure phase composition of the as-fabricated samples. The lattice parameters did not change by low-temperature oxidation at 530°C in an oxygen flow at 10 Pa amounting to $a°=°b=°0.556(2)°nm$, $c°=°0.789(3)°nm$.

The degree of superstructural Fe/Mo cation ordering [16] was estimated as $P°= 0.88$. On the other hand, traces of $SrMoO_4$ not exceeding 5.4% appeared in sample SFMO-III oxidized for 5 h. The average grain size derived from SEM pictures was 75 nm (Fig. 2).

An increase of the ion valence by further oxidation increases the binding energy and, thus, changes the peak position in the XPS spectrum since the electrons are now more tightly bound by additional nuclear charge. For this reason, we use XPS as a tool to study the ongoing Mo oxidation (Fig. 3). Upon oxidation, the envelope of the Mo-3d$_{3/2}$ and Mo-3d$_{5/2}$ peaks shift in opposite directions closer to the corresponding peaks of $SrMoO_4$ at 232.7 eV for Mo-3d$_{5/2}$ and 235.8 eV for Mo-3d$_{3/2}$ [23]. Deconvolution of the Mo-3d XPS spectra into $Mo^{5+}$ and $Mo^{6+}$ contributions was performed as described in detail elsewhere [24]. The distance between the doublets of the Mo-3d core-level spectra was set to 3.13 eV while the Mo-$3d_{5/2}$/Mo-$3d_{3/2}$ peak ratio amounted to 3/2 [25]. Following [26], the smaller peaks at 230.9 and 234 eV were attributed to the $Mo^{5+}$ state. The analysis of the peak areas reveals then 29% $Mo^{5+}$ and 71% $Mo^{6+}$ contributions to the Mo-3d spectrum of the as-fabricated sample SFMO-I in good agreement with data in the literature [26]. With increasing annealing time, an increasing fraction of $Mo^{6+}$ occurs giving evidence of a valence transition $Mo^{5+} \rightarrow Mo^{6+}$ of a part of the Mo ions which we relate to $SrMoO_4$ formation at the surface of the nanograins (Table 1). Note that XPS is a surface-sensitive technique with an average measurement depth of about 5 nm. Therefore, our results are similar to the ones obtained for the formation of a native $SrMoO_4$ surface barrier on air-exposed SFMO thin films [27].

**Table 1** – Mo-$3d_{5/2}$ and -$3d_{3/2}$ binding energies and fractions of $Mo^{6+}$ and $Mo^{5+}$ ions.

|  | Mo$3d_{5/2}$ | | Mo$3d_{3/2}$ | |
| --- | --- | --- | --- | --- |
|  | $Mo^{6+}$, eV (%) | $Mo^{5+}$, eV (%) | $Mo^{6+}$, eV (%) | $Mo^{5+}$, eV (%) |
| SFMO-I | 232.61 (71%) | 230.68 (29%) | 235.73 (71%) | 234.61 (29%) |
| SFMO-II | 232.47 (78%) | 230.81 (22%) | 235.68 (78%) | 234.53 (22%) |
| SFMO-III | 232,49 (86%) | 230.22 (14%) | 235.65 (86%) | 234.81 (14%) |

A recovery of the oxidized samples by annealing at 900°C in a 5%$H_2$/Ar flow returns the $Mo^{5+}$ fraction to the as-fabricated state.

An oxidation process localized at the grain boundaries affects the resistivity but does not influence the magnetization [8]. Therefore, we have studied the magnetization of the unannealed pressed and annealed samples. Fig. 4 illustrates the temperature dependence of the magnetization of samples SFMO-I and SFMO-III. Both samples were ferromagnetic with a Curie temperature of about 424 K. The low-temperature annealing did not provoke any appreciable changes in the magnetic properties giving evidence of the lack of substantial modifications in the grain bulk.

The assumption of nanograin surface oxidation was further proved by measurements of the electrical resistance. Fig. 5 shows the temperature dependences of the resistivity of the samples. The unannealed pressed sample SFMO-I exhibits a metallic conductivity at higher temperatures, an upturn point at about 35 K and a semiconductor-like conductivity below this upturn point (Fig. 5a). After annealing for 3 h in an argon flow, the upturn point of sample SFMO-II shifted to about 200 K. Finally, after a 5 h annealing, the metallic conductivity disappeared accompanied by an increase of the resistivity by about six orders of magnitude (Fig. 5b).

The resistivity behavior of sample SFMO-III is well described by the FIT model up to about 220 K. Table 2 compares the fitting parameters of Eq. (2.5) with FIT model data available for SFMO ceramics in literature [7] as well as with our fits using data of previous works on SFMO ceramics subjected to different sintering procedures [1,2,28,29]. It illustrates the impact of sample processing on both the barrier height and width.

We attribute the relatively low values of $T_1$ and $T_0$ of our sample compared to the ones fabricated under similar conditions [2] to a smaller barrier area $A$. This is qualitatively consistent with an increased value of $\rho(0)$. A further increase of $\rho(0)$ is caused by a more than three orders of magnitude higher density of barriers in our nanograined material.

**Table 2** – FIT model parameters of SFMO ceramics. *d* is the averaged grain size, *P* is the superstructural ordering parameter of Fe and Mo cations at B-sites.

| Sample | $\rho(0)$, $\Omega\cdot$cm | $T_1$, K | $T_1/T_0$ | Ref. |
|---|---|---|---|---|
| SFMO-I, $d$ = 74nm, $P$ = 0.88 | $3.39\cdot10^5$ | 450 | 1.78 | this work |
| sintered SFMO, $d$ = 2μm | $1.34\cdot10^{-2}$ | 5171 | 2.09 | [7] |
| cold-pressed SFMO, $d$ = 1μm | 1.39 | 1586 | 2.73 | [7] |
| reduced SFMO | $2.96\cdot10^{-2}$ | 6989 | 2.74 | [7] |
| SFMO sintered for 2h at 1200°C in 1%H$_2$/Ar, $P$ = 0.87 | 37.2 | 25.6 | 0.18 | [1] |
| SFMO sintered at 1100°C in vacuum | 69.3 | 464 | 1.14 | [28] |
| SFMO long-term (16 h) sintered at 1200°C in 5%H$_2$/Ar, $P$ = 0.94 | 2.38 | 2358 | 2.14 | [2] |

The temperature of 220 K, at which the thermal energy *kT* overcomes the energy barrier, gives an estimate of the barrier hight yielding $V_0 \sim 20$ meV.

According to Eq. (2.3), the ratio $T_1/T_0$ determines the barrier width *w*. Considering in Eq. (2.4) an effective mass of $m_{eff}$ = 2.5 $m_e$ [30], we obtain for $T_1/T_0$ = 1.78 a barrier width of *w* = 1.24 nm. This is equal to the size of about three stacked BO$_6$ octahedra, where B are either Fe or Mo ions. The estimated barrier width yields a barrier area of $A \approx 300$ nm$^2$, which is reasonable for nanograins with an average size of 75 nm.

An independent estimation of the barrier width arises from the breakdown voltage amounting to 4 V for similar dielectric SrMoO$_4$ shells covering SFMO nanoparticles [18]. The occurrence of the electrical breakdown in metal oxide dielectrics is determined by the local electric field and the chemical bond strength. Therefore, only a certain energy density equivalent to a limit in electrostatic pressure may be stored in a dielectric. This results in an empirical relationship [31]:

$$E_{bd} \approx 2450\cdot\varepsilon^{-1/2} \text{V}/\mu\text{m}, \qquad (4.1)$$

where $\varepsilon$ is the relative dielectric permittivity. The $\varepsilon$ values of SrMoO$_4$ range from about 10 for nanopowders synthesized by combustion method [32] to an average of about 35 for single crystals [33]. The corresponding thicknesses of the nanoparticle surface layer, which represent a measure of the barrier width, range then from 2.6 to 4.8 nm in good agreement with a SEM image of similar nanosized SFMO ceramics annealed at 430°C in an oxygen flow [22].

Near room temperature, the Mott variable-range hopping [34,35] appears described by

$$\rho = \rho_0' \cdot \exp\left(\frac{T_0'}{T}\right)^{1/4} \tag{4.2}$$

with

$$\rho_0'(T) = \frac{\sqrt{8\pi k \chi'/n(E_F)}}{2e^2 \nu_{ph}} \cdot T^{-1/2}, \tag{4.3}$$

and

$$kT_0' = \frac{18\chi'^3}{n(E_F)}, \tag{4.4}$$

where $n(E_F)$ is the electronic density of states at the Fermi level in the absence of electron-electron interactions, $T_0'$ is a characteristic temperature being $T_0' = 3.77\times10^6$ K for sample SFMO-III, and $\nu_{ph} = k\cdot\Theta_D/h$ is the phonon frequency at the Debye temperature, $\Theta_D = 338$ K [36]. The $T_0'$ value varies between $T_0' = 2.02\times10^6$ K, obtained for similar nanograined (grain size 75 nm) SFMO ceramics finally annealed in a 5% H$_2$/Ar flow at 950°C for 4 h with a parameter of superstructural Fe/Mo cation ordering of $P = 0.88$ [22], and $T_0' = 3.55\times10^7$ K derived for SFMO ceramics fabricated by microwave sintering of SrCO$_3$, Fe$_2$O$_3$ and MoO$_3$ as raw materials at 10 Pa in the presence of granular activated carbon with $P = 0.063$ [37]. Polycrystalline SFMO samples prepared by sol-gel method and finally sintered for 3 h at 1200°C in a reducing atmosphere provided by active carbon particles, possess a characteristic temperature of $T_0' = 9.13 \times 10^6$ K [38].

The Mott variable range hopping mechanism is characterized by a hopping activation energy

$$E_h = \frac{1}{4} \cdot kT^{3/4} T_0'^{1/4}, \tag{4.5}$$

amounting to 72 meV at 300 K.

The application of a magnetic field increases the probability of electron tunneling through dielectric barriers decreasing the resistivity and recovering the metallic conductivity of the nanograins [1,39]. For instance, the upturn temperature decreases from 180 K at 0.2 T to about 40 K at 7 T, thus providing a sufficient low-field magnetoresistance promising for device applications [1].

## 5. Conclusions

In this work we have demonstrated the controlled formation of dielectric SrMoO$_4$ barriers between nanograins in SFMO ceramics upon annealing. To elucidate the process, measurements of the XPS, XRD, and temperature dependences of the magnetization and electrical resistivity after each annealing step were performed. The barriers substantially increase the resistivity of the SFMO ceramics. In annealed samples, the conductivity occurs via a fluctuation-induced tunneling between the grains, and the metallic behavior of the conductivity disappears. At higher temperatures, Mott's variable-range hopping conductivity mechanism was identified, and the hopping activation energy was estimated.

**Acknowledgements**

This work was developed within the scope of the European project H2020-MSCA-RISE-2017-778308–SPINMULTIFILM. At the University of Aveiro, the work was partially supported by the project i3N, UIDB/50025/2020 & UIDP/50025/2020, financed by national funds through the FCT/MEC.

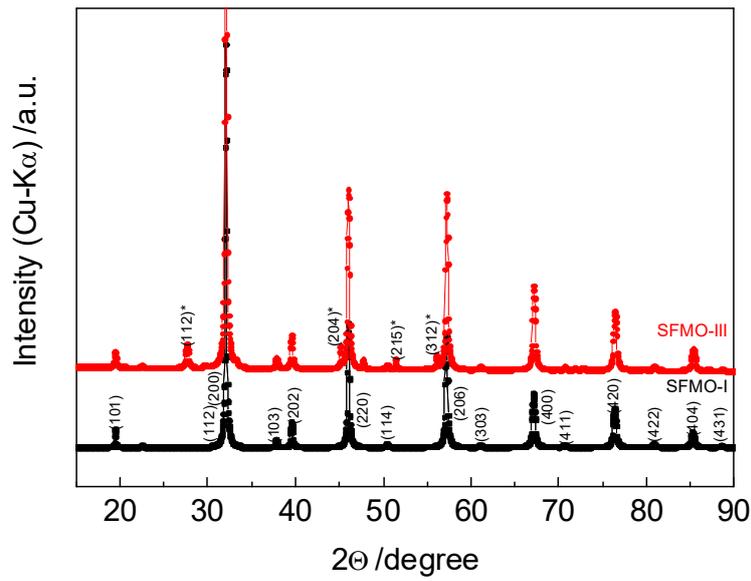

**Fig. 1** – XRD patterns of sample SFMO-I (unannealed pressed powder) and of sample SFMO-III annealed at 530°C in an Ar flow for 5 h. The $SrMoO_4$ peaks are marked by asterisks.

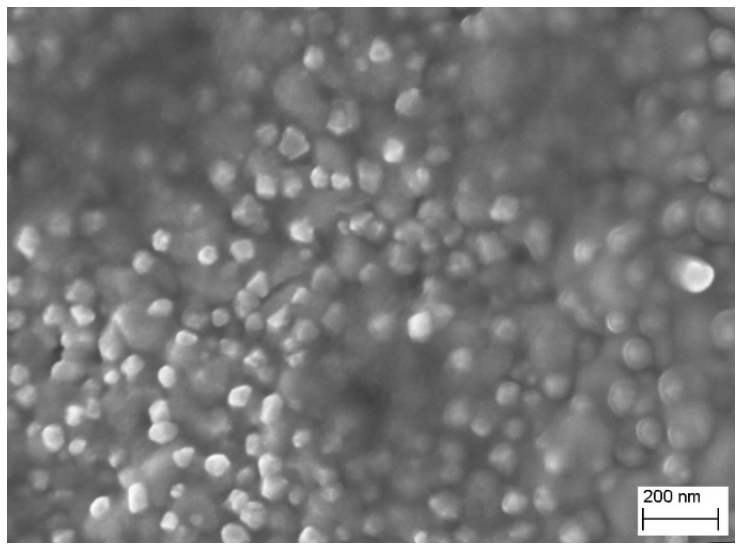

**Fig. 2** – A SEM micrograph of the microstructure of the SFMO powder deposited by centrifugation onto a corundum ceramics substrate. The average grain size was estimated to be about 75 nm.

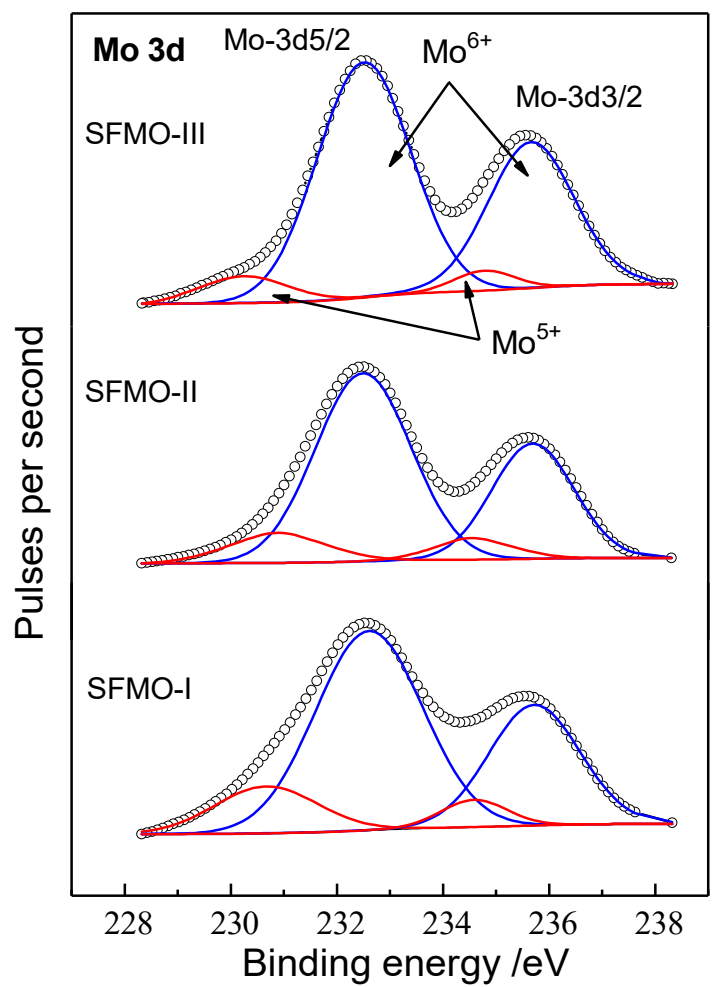

**Fig. 3** – XPS spectra of the Mo-3d core levels of samples SFMO-I, SFMO-II, and SFMO-III. Open circles represent experimental spectra, red solid lines represent the $Mo^{5+}$, the blue ones correspond to the $Mo^{6+}$ valence states.

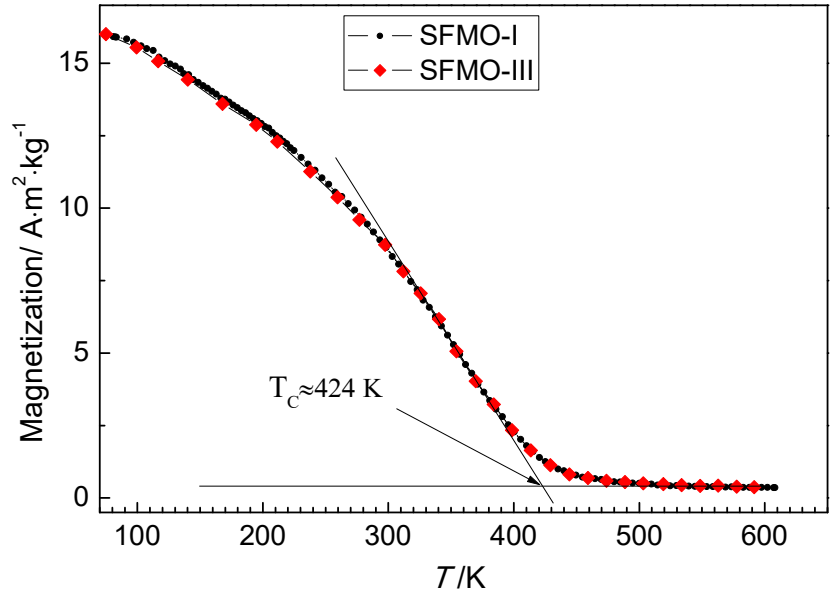

**Fig. 4** – Temperature dependences of the magnetization of the unannealed pressed powder sample SFMO-I and of that annealed at 530°C in an Ar flow for 5 h – SFMO-III. The magnetic measurements were carried out in a magnetic field of $B = 0.86$ T.

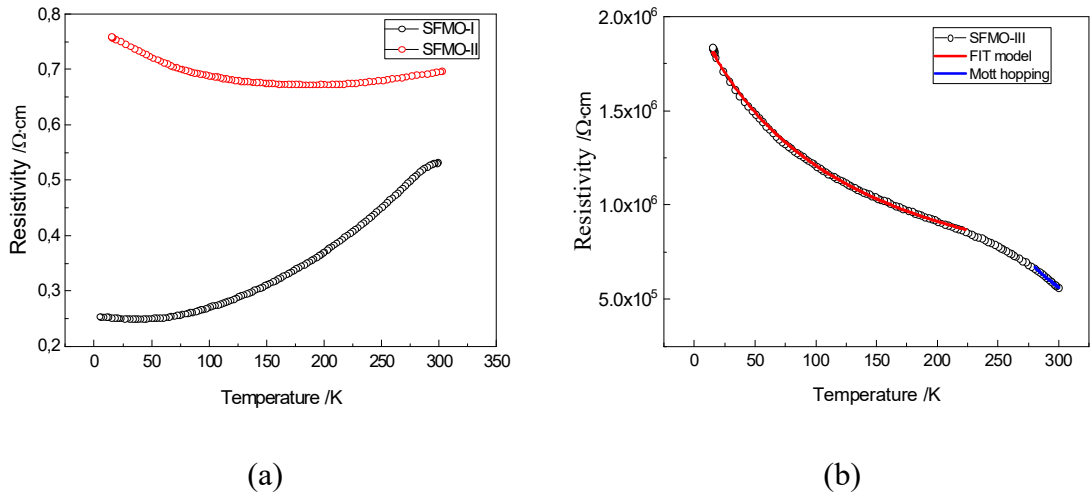

(a)　　　　　　　　　　　　　　(b)

**Fig. 5** – Temperature dependence of the resistivity of samples SFMO-I, SFMO-II (a), and SFMO-III (b).